\documentclass{article}
\usepackage[T1]{fontenc}
\usepackage[utf8]{inputenc}
\usepackage[]{ismir} 
\usepackage{amsmath,cite,url}
\usepackage{graphicx}
\usepackage{color}
\usepackage{booktabs}
\usepackage{array}
\usepackage{amssymb}

\title{Ghost in the Encoder: Decodable Artist Identity Representations in Lyrics-to-Song Generation}

\multauthor
  {Arhan Vohra* \hspace{1cm} Choenden Kyirong* \hspace{1cm} Laura Ibáñez-Martínez \hspace{1cm} Martín Rocamora}
  {
  Music Technology Group, Universitat Pompeu Fabra\\
  {\tt\small arhan.vohra01@estudiant.upf.edu, choendenkyirong@gmail.com}\\
  \small\textit{*Equal contribution}
  }

\def\authorname{A. Vohra, C. Kyirong, L. Ibáñez-Martínez, M. Rocamora}

\begin{document}

\maketitle

\begin{abstract}
Text-to-song generation models can be prompted to imitate specific artists or regurgitate entire songs from their training data. Although these phenomena have been documented behaviorally on small datasets, little is known about the internal representations that may give rise to them. Prior interpretability work on generative audio has focused on locating semantic concepts such as genre or time signature within model activations. In this work, we show that a trained model can be probed for linearly decodable representations of artist identity from song lyrics alone, without any additional identifiers. Through a controlled case study of ACE-Step 1.5 spanning 2,000 songs across 100 artists, we demonstrate that the artist associated with a given set of lyrics can be identified within the model’s internal activations, and that this conditioning signal propagates from the lyric encoder to the diffusion backbone during inference. These findings indicate that lyrics constitute an artist-level conditioning channel not addressed by prompt-side replication safeguards. More broadly, our work highlights how latent-space analysis can be used to audit what generative music models have implicitly learned from their training data.
\end{abstract}

\section{Introduction}\label{sec:introduction}

Consider the challenge of teaching music students to appreciate the works of Bob Dylan. Perhaps we'd like for them to learn from Dylan how to craft songs imaginatively across many genres while preserving their voice as songwriters. Unfortunately, some students may just memorize every lyric from ``Highway 61 Revisited'', and others might incorporate crude impressions of Dylan's vocal delivery and harmonica solos in their own songs. The difficulty is that all of these outcomes require exposure to the same body of work, and the line between \textit{generalization}, \textit{memorization}, and \textit{impersonation} is drawn not by what was heard, but by what was learned from it. 

Recent studies have shown that AI music generation models exhibit all three of these behaviors to varying extents: song-level memorization \cite{roh2025bobs, yuan2025yue}, artist-level impersonation \cite{coelho2025artist, nagarajan2025namefree}, and concept-level generalization of features such as genre and mood \cite{singh2026discovering, staniszewski2026tada, koo2025smitin, wei2024music}.

While generative models can be probed mechanistically to reveal interpretable representations of musical semantics (e.g.,\ time signature) and specific genres, their tendency to replicate artist-level characteristics has only been documented at the behavioral level. Coelho~\cite{coelho2025artist} presented descriptive evidence that closed-source models can be conditioned to imitate certain artists via metatag-informed prompting, while Nagarajan \& Dong \cite{nagarajan2025namefree} used LLM-generated descriptors and CLAP-based similarity to test artist conditioning on a limited two-artist dataset with MusicGen. Roh et al. \cite{roh2025bobs} showed that full-song replication can also be elicited by prompting models with the lyrics from popular songs, and that this behavior is preserved after homophonic replacement of the song's lexical content (``Bob's confetti'' from ``mom's spaghetti''). However, song generation is approached as a black-box process in all three of these accounts, and evaluation has been limited to fewer than 50 songs.

Our study directly addresses the need for \emph{latent-space} explanations expressed in prior work \cite{roh2025bobs} by proposing a mechanistic approach to understanding how music generation models develop artist-specific representations from lyrics alone, using linear probing \cite{alain2017understanding} across the conditioning pipeline. We demonstrate its effectiveness with a systematic case study on ACE-Step 1.5 \cite{gong2026acestep15}, a state-of-the-art lyrics-to-song model with open weights and relatively modest compute requirements. Our class-balanced dataset spans 2,000 songs from 100 artists across 5 genres, namely: pop, rock, hip-hop, R\&B, and country. To the best of our knowledge, this is the first interpretability study of artist representations in music generation.

We show that the lyric encoder in ACE-Step 1.5 produces highly decodable linear representations of artist identity without receiving any explicit artist information in its input. We define \textit{artist identity} as the artist label associated with a song's lyrics. This is a functional definition rather than a perceptual feature of music, allowing us to show that artist-identifying representations exist within a music model without making assumptions about their relationship to the acoustic output. 

The representations remain linearly decodable through the model's cross-attention layers into the denoising process and in the generated audio itself. We examine how decodability varies with artist genre, lexical identifiers, as well as the text-conditioning pipeline, and consider how our findings may inform future work on artist-level representations.

\section{Related Work}\label{sec:related_work}

\subsection{Replication and Memorization in GenAI}
Memorization and replication are well documented phenomena in generative models across modalities. Language models have been shown to leak verbatim text sequences from their training data in response to prompt-based attacks, including private information about individuals \cite{carlini2021extracting} and copyrighted books in their entirety \cite{nasr2025scalable}. Image generation models are similarly capable of producing near-perfect replicas of their training data \cite{carlini2023extracting}. This behavior is known to be amplified by the presence of duplicate entries in the training set \cite{somepalli2023understanding}, and can be triggered via ``highly specific captions'' seen during training \cite{Somepalli2022DiffusionAO}.

This propensity for generative models to regurgitate their training data has also been observed in music models. Some open-source model releases, such as YuE \cite{yuan2025yue} and MusicGen \cite{copet2023musicgen}, include memorization studies to quantify this phenomenon, and ACE-Step warns users of the potential for ``unintentional copyright infringement'' on its GitHub page \cite{acestep15_readme}. Recently, Roh et al.~\cite{roh2025bobs} established that song replication can be triggered by prompting YuE with verbatim song lyrics, and that prompt-side guardrails to prevent this in closed-source models like Suno can be bypassed by modifying the lyrics with semantically different text sequences that maintain the original's phonetic structure.

Other studies have explored the use of descriptive prompts to elicit artist imitation without explicitly mentioning their names, in order to circumvent prompt filtering. Coelho \cite{coelho2025artist} proposed the use of tags sourced from publicly available metadata to curate list-like artist prompts, with anecdotal evidence that this reproduces their idiosyncratic stylistic choices during generation. Additionally, Nagarajan \& Dong \cite{nagarajan2025namefree} described a process to generate policy-compliant artist prompts using LLMs, and used audio embeddings to compare the generated audio against a reference track. 

While a broader body of work has addressed training data attribution and anti-memorization interventions in generative audio models \cite{barnett2024exploringmusicalrootsapplying, choi2025largescaletrainingdataattribution, messina2026mitigatingdatareplicationtexttoaudio, DBLP:conf/icassp/BraliosWGPKHR24}, the mechanisms that may enable artist-level imitation and song-level replication have not yet been dissected.

\subsection{Lyrics-to-song Models}
Current lyrics-to-song systems are composed of several components with distinct responsibilities, including pretrained text encoders, audio codecs, alignment modules, and generative backbones \cite{yuan2025yue, liu2025songgen, ning2025diffrhythm, gong2025acestep}. The generative backbones can follow autoregressive \cite{dhariwal2020jukebox, yuan2025yue, liu2025songgen}, diffusion-based \cite{ning2025diffrhythm, evans2024stable}, or hybrid \cite{gong2025acestep, yang2025songbloom} paradigms. Across all three, text content enters the generative process through pretrained encoders that may be jointly fine-tuned with the rest of the model.

We study ACE-Step 1.5 \cite{gong2026acestep15}, a hybrid model in which a Qwen-based language model generates structured Chain-of-Thought metadata while a Diffusion Transformer (DiT) synthesizes audio. The DiT receives conditioning through three parallel pathways merged via cross-attention: a caption encoder (Qwen3-0.6B) \cite{yang2025qwen3technicalreport}, a timbre encoder (if reference audio is provided), and a lyric encoder based on an 8-layer transformer and Qwen3 token embeddings. Notably, ACE-Step 1.5 generates songs significantly faster than other open-weight models of comparable quality, making it a particularly feasible choice for our case study.
\subsection{Interpretability of Music Generation Models}

The linear representation hypothesis \cite{park2024linear} posits that high-level concepts can be captured by linear directions in a vector space, and probing the hidden layers of a model \cite{alain2017understanding} can reveal which concepts it has internalized. Further, these can be used to quantify and prevent adverse behavior such as gender bias in language models \cite{vig_investigating_2020}. 

Wei et al.~\cite{wei2024music} probed MusicGen for music-theoretical concepts, finding that certain layers encode these properties more strongly than others, while SMITIN \cite{koo2025smitin} also leveraged probing to identify and steer attention heads responsible for traits such as instrument presence. Most recently, Singh et al.~\cite{singh2026discovering} discovered interpretable features from MusicGen's residual stream with sparse autoencoders, and TADA ~\cite{staniszewski2026tada} demonstrated that musical concept representations can be localized to specific layers of diffusion models.

Our study draws from this line of inquiry to investigate how text-based conditioning relates to artist identity representations in ACE-Step 1.5, and whether this relationship can functionally explain the imitation and replication behavior observed in prior work.

\section{Method}\label{sec:method}

\subsection{Dataset}
We construct our prompting dataset by retrieving English-language song lyrics from the top 100 artists equally distributed across 5 genres (rock, pop, hip-hop, R\&B, and country) on Last.fm, as of April 2026. For each artist, we fetch their top 20 songs from the Genius.com API (ranked by the API's internal popularity metric), excluding variations (remixes, covers, edits) and songs released after 2024. We anonymize any appearances of a given artist's own name in their lyrics by using the phonetic replacement method described in \cite{roh2025bobs}. Since ACE-Step 1.5's ~1.8M-song training corpus is undocumented, training-set membership cannot be verified directly; our selection of popular pre-2024 songs is intended to maximize its likelihood.

\subsection{Prompting and Activation Extraction}

To observe how ACE-Step encodes and propagates artist information during inference, we insert \texttt{pytorch} hooks into intermediate layers while generating songs conditioned on the datasets' lyrics, storing the activations for probing.

The ACE-Step 1.5 generation pipeline contains encoders for several conditioning pathways (text caption, lyrics, timbre transfer) which are merged together into the DiT's cross-attention mechanism. This model also offers a unique conditioning pathway through its chain-of-thought (CoT) module, which can be used to generate a relevant song caption, as well as tempo and key conditioning, based on a lyric prompt alone. In order to understand how the CoT process modifies artist encoding, we consider two configurations of the inference process: \texttt{no-CoT} (where the reasoning model is deactivated) and \texttt{with-CoT}. The key differences are shown in Table \ref{tab:inference-configs}.

All experiments use \texttt{acestep-v15-turbo}, called with the following inputs for consistency: \texttt{duration=180s, bpm=None, key=None, text\_caption=None, timbre\_reference=None}. In the \texttt{with-CoT} configuration, the tempo, key, and text caption inputs are overridden by the CoT model.

In both configurations, we pass in song lyrics from the dataset and extract activation tensors from the following locations:

\textbf{Conditioning stream outputs $(\mathbb{R}^{2048})$:} token-level outputs of the lyric encoder, text/caption stream, (unused) timbre transfer stream, and the merged conditioning sequence passed to the DiT.

\textbf{Lyric encoder hidden layers $(\mathbb{R}^{2048})$:} token-level hidden states from the 8 lyric-encoder transformer layers.

\textbf{DiT cross-attention K/V projections $(\mathbb{R}^{1024})$:} key and value projections of the merged conditioning sequence, extracted at 5 equally-spaced pairs of DiT layers (10 layers total). This accounts for the model's hybrid architecture, where alternating layers use sliding-window (local) or group-query (global) attention.

\textbf{DiT audio-latent states $(\mathbb{R}^{2048})$:} pre-cross-attention states, cross-attention outputs, and post-cross-attention states, extracted at the same 10 DiT layers across 8 denoising timesteps.

Together, these probes trace artist-identifying information from lyric encoding, through merged conditioning and cross-attention, into the DiT denoising trajectory.

\begin{table}[t]
\centering
\small
\renewcommand{\arraystretch}{1.2}
\setlength{\tabcolsep}{6pt}
\begin{tabular}{@{} l c c @{}}
\hline
\textbf{Conditioning input} & \texttt{no-CoT} & \texttt{with-CoT} \\
\hline
Lyric encoder & lyrics & lyrics \\
Caption       & \textit{null} & generated from lyrics \\
BPM / key     & \textit{null} & generated from lyrics \\
Timbre        & \textit{null} & \textit{null} \\
\hline
\end{tabular}
\caption{Conditioning inputs to the DiT under the two inference configurations. Both pass the same lyrics to the lyric encoder; \texttt{with-CoT} additionally activates the chain-of-thought module, which derives a caption and infers BPM/key. All other inputs are held fixed.}
\label{tab:inference-configs}
\end{table}

\subsection{Internal Probing}

For each hook point $h$, we quantify the extent to which artist identity is linearly decodable \cite{alain2017understanding}. We measure this using the accuracy of a multi-class classifier trained on $h$.

Let $\{(x_i, y_i)\}_{i=1}^N$ be a dataset, where $x_i$ is an input example and $y_i \in \mathcal{Y}$ is the corresponding label. Let $H_h(x_i)$ denote the activation tensor captured at hook point $h$.

Since activations may have non-feature axes (e.g., time or token dimensions), we convert each activation tensor into a single feature vector by averaging over all non-feature axes. This yields
\[
a_i = \operatorname{mean}(H_h(x_i)) \in \mathbb{R}^{d_h},
\]
where the mean is taken over all axes except the final feature dimension. For example, if $H_h(x_i) \in \mathbb{R}^{T \times d_h}$, then
\[
a_i = \frac{1}{T} \sum_{t=1}^{T} H_h(x_i)_t.
\]

For each cross-validation fold $k$, we train a linear multi-class logistic regression classifier on the corresponding training split $T_k$:
\[
p_{\phi^{(k)}}(y \mid a_i)
= \operatorname{softmax}\!\left(W^{(k)} a_i + b^{(k)}\right)
\]
where $\phi^{(k)} = \{W^{(k)}, b^{(k)}\}$. The training objective for each fold is the cross-entropy loss evaluated on $T_k$. 

\subsection{Output-side Probing}

While probing can establish that intermediate activations contain linear representations of artist identity, we need a scalable metric to detect artist identity in the generated audio. We use LAION-CLAP and MuQ-MuLan embeddings \cite{wu2024largescalecontrastivelanguageaudiopretraining, zhu2025muqselfsupervisedmusicrepresentation} to analyze whether artist identity is decodable from the generated output. This approach is well-supported by previous studies on concept-based interpretability and similarity in audio, which use CLAP-like models as a quantitative proxy to measure semantic or musical alignment \cite{singh2026discovering, koo2025smitin, staniszewski2026tada}, and recently, artist similarity in music generation \cite{nagarajan2025namefree}. Mirroring the probing approach described earlier, we extract one embedding per generated song and train a linear classifier on the embeddings to test whether artist information remains detectable in the audio output.

Additionally, we compare ACE-Step's generated audio to the corresponding original recordings with song-level similarity scores from CLEWS-DVI \cite{serra2025supervisedcontrastivelearningweaklylabeled}, a version-matching embedding model, in order to understand whether artist decodability may be related to song-level replication. This follows the use of version-matching models in systematic studies of replication \cite{yuan2025yue}.

\section{Experiments}\label{sec:experiments}
\subsection{Probing the Conditioning Encoders}

\begin{figure}
    \centering
    \includegraphics[width=\linewidth]{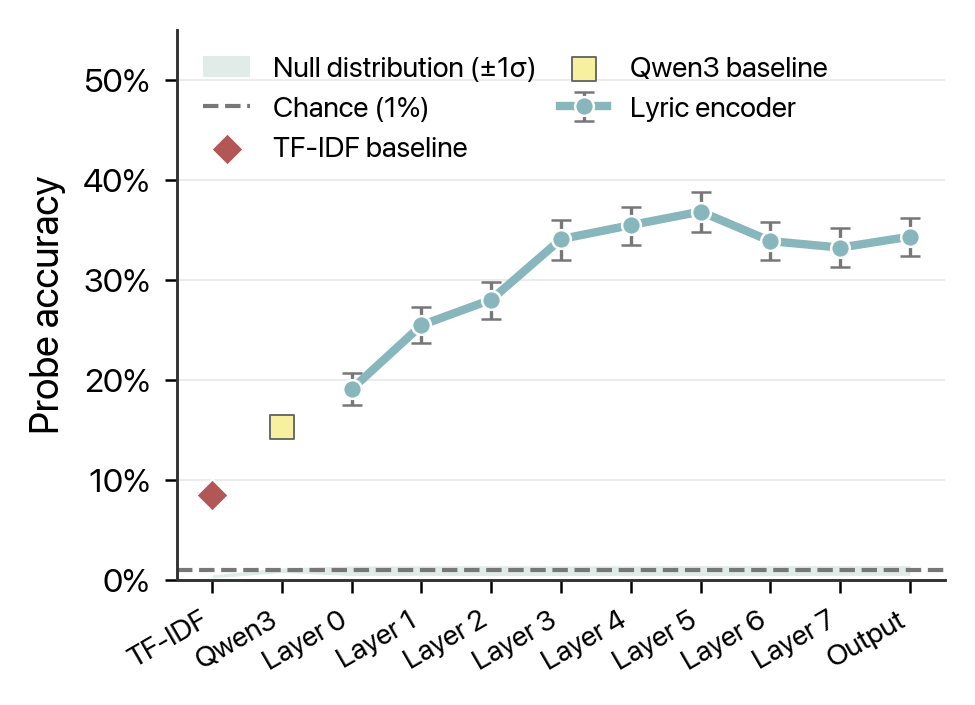}
    \caption{Artist probe accuracy rises through the lyric encoder layers. Error bars show the 95\% confidence interval.}
    \label{fig:rq1}
\end{figure}

\textbf{RQ1: Does the lyric encoder develop linearly decodable representations of artist identity?}

We prompt ACE-Step 1.5 with song lyrics over our entire 2,000 song dataset and extract activations from each layer of the lyric encoder as previously described. We report held-out accuracy from the probes after 5-fold cross validation and permutation tests $(n=1000)$. These results are compared against identically structured probing experiments on two baselines: (1) the pre-trained Qwen3 token embeddings used as the encoder's input, and (2) bag-of-words (BoW) based on TF-IDF.

A probe trained on the pooled output of the lyric encoder correctly identifies the source artist with an accuracy of 0.344, approximately 4 times the BoW baseline (0.085), 35 times higher than random chance, and well outside the null distribution $(p<0.001)$.

Artist-discriminative representations rise through the first five layers of the encoder and plateau thereafter, peaking at layer 5 (0.361) before stabilizing between 0.34–0.36 in layers 5–7. The pre-trained Qwen3 embedding produces an artist probe accuracy of 0.153; although artist decodability at the encoder's intake is already better-than-random, the learned layers amplify this substantially (Figure \ref{fig:rq1}).

We also find that genre identity is decodable from the lyric encoder output with an accuracy of 0.764, raising the question of whether the artist probe is detecting meaningful artist-specific signals or is confounded by genre-level representations.

\textbf{RQ2: Is the probe detecting artist or genre identity?}

\begin{figure}
    \centering
    \includegraphics[width=\linewidth]{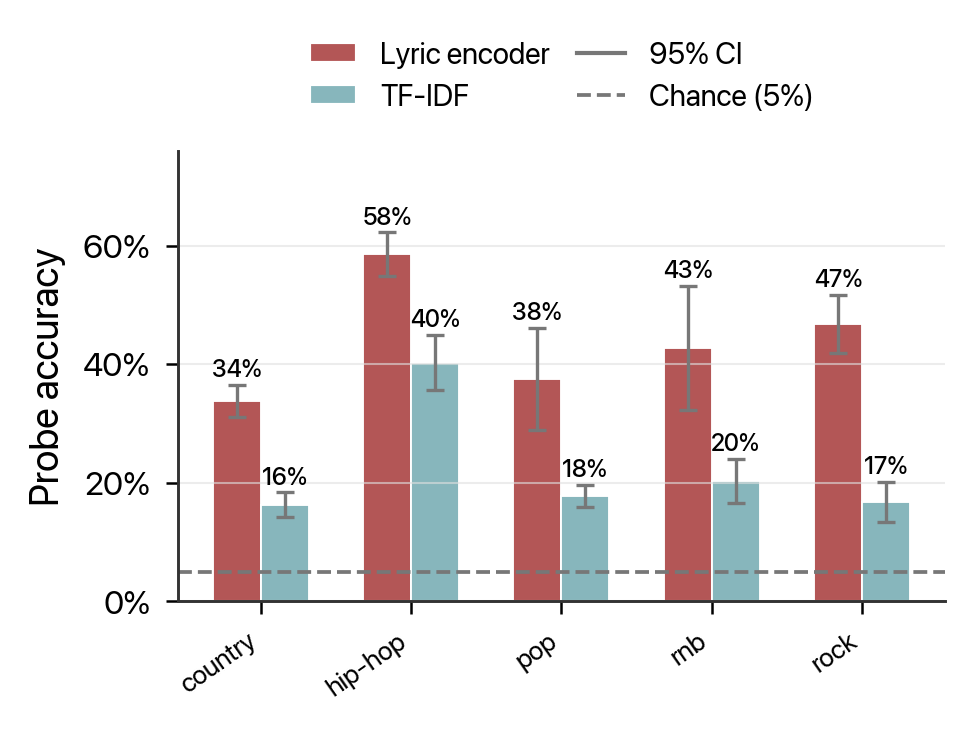}
    \caption{Genre-wise mean artist decodability from the lyric encoder's output.}
    \label{fig:rq2}
\end{figure}

To rule out genre as a confounding factor, we probe artist identity within each genre subset of our data. This tests whether the probe captures individual artists or merely the vocabularies and songwriting norms of their genres.

Figure \ref{fig:rq2} shows that mean artist decodability at the encoder varies substantially across genres, from 0.585 for hip-hop down to 0.338 for country. Hip-hop is also the genre where the TF-IDF baseline is strongest (0.403), suggesting that lexical distinctiveness between artists accounts for a larger share of the signal here than in other genres, where probing the encoder is over twice as accurate as bag-of-words.

\subsection{Probing Downstream Generation}

\textbf{RQ3: Does the DiT preserve artist-level decodability during generation?}

\begin{figure*}
    \centering
    \includegraphics[width=\textwidth]{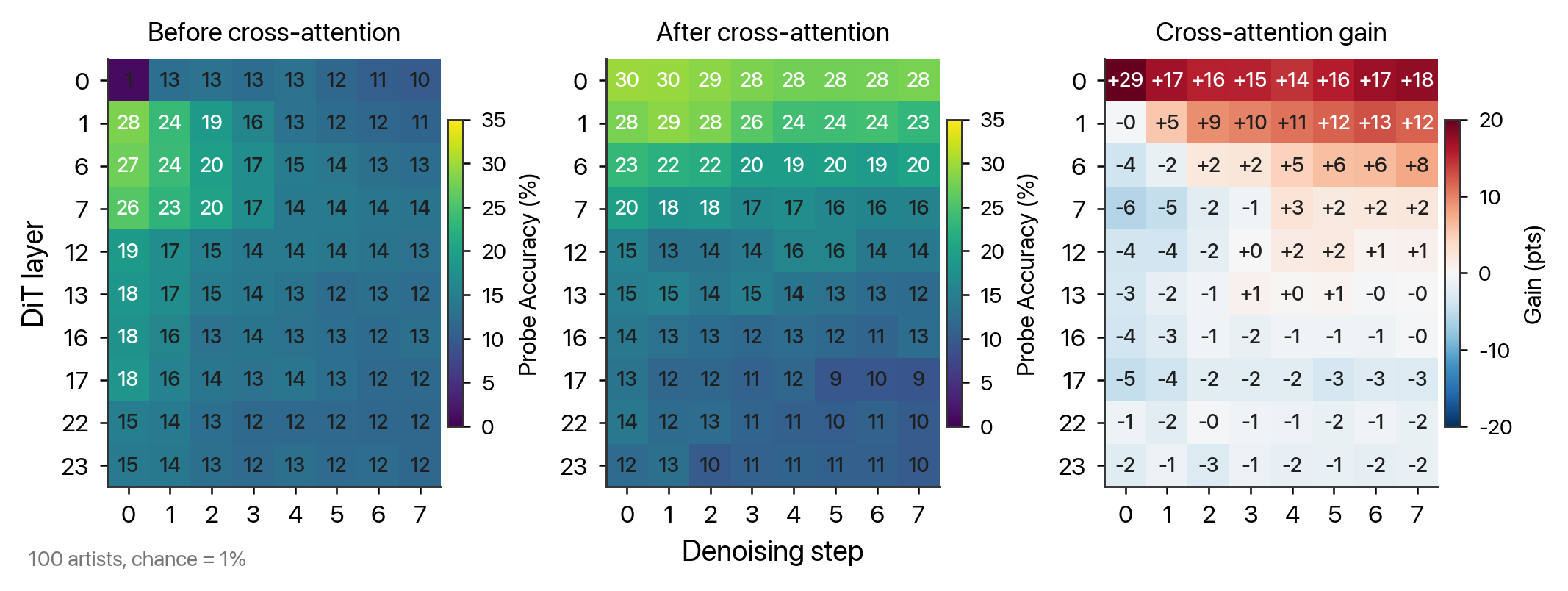}
    \caption{Cross-attention injects decodable artist representations into the first few layers of the DiT audio-token hidden states; these become progressively less decodable in deeper layers.}
    \label{fig:rq4}
\end{figure*}

To understand if the Diffusion Transformer actually retains the learned artist representations, we probe the cross-attention K/V projections and audio-token hidden states during denoising. The generator does not receive any style caption or metadata (\texttt{no-CoT}).

Our probe reveals that artist identity remains strongly decodable, albeit slightly attenuated by the cross-attention interface, at 0.292 on average at the K/V projections (a 15\% decrease from 0.344 at the lyric encoder). Individual artist decodability is highly correlated between the lyric encoder and cross-attention probes $(\rho \approx 0.90, r \approx 0.91)$, suggesting that the most decodable artists are similar at both points.

In the audio-token hidden states (Figure \ref{fig:rq4}), we see an emerging pattern in how the model attends to these representations during the diffusion process. The noisy latent in layer 0 has a probe accuracy of exactly 0.01 before denoising, matching random chance classification between 100 artists. In the first layers, especially layers 0 and 1, artist decodability sharply increases before and after being re-weighted by cross-attention. From layers 12-13, artist decodability is generally lower (0.10-0.14 in the last layers), with gradual decay over consecutive time steps.

Prior work has suggested that concept representations can be localized to specific layers of generative music models \cite{staniszewski2026tada}, and that diffusion models may internalize or ``lock in'' different concepts at different steps in the denoising process \cite{gorgun2026temporalconceptdynamicsdiffusion}. The early layers may be re-encoding artist information into acoustic structure, developing non-linear representations, or discarding it entirely. While a causal evaluation of these representations would be required to positively distinguish amongst these possibilities, we examine the model's audio outputs in our final experiments (RQ4-5) as an end-to-end behavioral check on what survived the generation process.

\textbf{RQ4: Is artist identity detectable in the generated audio?}

Our internal probes establish that artist identity is linearly decodable at successive stages of ACE-Step's conditioning pipeline, but this does not confirm that the signal reaches the generated output. We test this by extracting MuQ-MuLan and CLAP embeddings from the generated audio and train the same probing classifier on these embeddings. 

Across all 100 artists, the MuQ-MuLan probe achieves a mean accuracy of 0.109 under \texttt{no-CoT} conditioning. The same probe applied to original song recordings reaches 0.449, indicating that the generation process may preserve some artist-discriminative information. The CLAP-based probe shows considerably lower artist decodability (0.043) overall, which may relate to MuQ's music-specific pretraining.

In order to control for genre, we repeat the probe for each genre class (Table~\ref{tab:audio-probes}) showing that the artist signal survives genre controls in all five genres to different degrees, with MuQ-MuLan accuracy ranging from 0.128 (pop) to 0.208 (rock) at a chance level of 0.05. The genre ranking at the audio level partially diverges from the encoder: rock artists are more decodable than hip-hop artists on average. This suggests that the features driving encoder-level decodability may differ from those that survive into the generated audio. However, since  neither embedding model has access to the input lyrics, any artist information recovered by the probe must still be present in the acoustic properties of the generated audio itself due to the conditioning of the lyric encoder.

These results confirm that artist-specific conditioning persists through the full generation pipeline, from lyrics through internal representations to the acoustic properties of the generated audio, and that this signal operates at the artist level rather than reflecting song specific memorization alone.

\textbf{RQ5: How does CoT conditioning modify artist representations in the generation pipeline?}

The chain-of-thought LM is somewhat unique to ACE-Step 1.5: it is simultaneously responsible for optimizing and formatting the user's inputs, generating missing conditions (text caption, tempo, key), and subsequently producing a text embedding based on them, which is used as latent conditioning for the DiT. We broadly assess how this module may impact artist identity representations.

In our \texttt{no-CoT} experiments, the text caption embeddings probed artist identity at near-chance accuracy (0.008), due to this conditioning stream being inactive. When the LM is turned on, artist identity probes at 0.069 at the same point, which is higher, but still below our baseline TF-IDF lyric classifier (0.085). We also find that a TF-IDF classifier trained on the LM-generated caption has a comparable probing accuracy to the one trained on lyrics (0.073), suggesting that any caption decodability may be lexically-grounded and not necessarily learned from training.

\begin{table*}[t]
\centering
\begin{tabular}{@{}lccccccc@{}}
\toprule
& \multicolumn{3}{c}{MuQ-MuLan} & & \multicolumn{3}{c}{CLAP} \\
\cmidrule(r){2-4} \cmidrule(l){6-8}
& Originals & no-CoT & with-CoT & & Originals & no-CoT & with-CoT \\
\midrule
\textit{Full dataset (100 artists, chance = 0.01)} \\
All genres & 0.449 & 0.109 & 0.101 & & 0.224 & 0.043 & 0.042 \\
\midrule
\textit{Within-genre (20 artists, chance = 0.05)} \\
Hip-hop & 0.445 & 0.180 & 0.183 & & 0.273 & 0.115 & 0.130 \\
Rock    & 0.659 & 0.208 & 0.130 & & 0.394 & 0.108 & 0.073 \\
R\&B    & 0.510 & 0.165 & 0.140 & & 0.250 & 0.083 & 0.100 \\
Country & 0.558 & 0.145 & 0.128 & & 0.368 & 0.088 & 0.053 \\
Pop     & 0.413 & 0.128 & 0.100 & & 0.300 & 0.073 & 0.068 \\
\bottomrule
\end{tabular}
\caption{Artist probe accuracy on generated audio embeddings. All results $p < 0.001$.}
\label{tab:audio-probes}
\end{table*}

Since the lyric encoder is deterministic and works in parallel to the CoT module, probe accuracy here remains unchanged across both conditions. However, decodability drops from 0.344 at the lyric encoder to 0.085 at the merge point where the LM's embeddings are concatenated with the lyric encoder output, and consequently, the K/V projections (0.292 $\rightarrow$ 0.089). Still, this may reflect the dilution of the lyric stream's per-token signal under mean-pooling rather than a loss of artist information in the merged sequence.

We also compare the audio outputs produced under both conditions using CLAP/MuQ-MuLan classification and cover identification Retrieval$@K$ via CLEWS. The songs produced by \texttt{with-CoT} are harder to probe for artist identity than those from \texttt{no-CoT} (Table \ref{tab:audio-probes}), which would suggest that retained artist-level conditioning in the audio is weaker when an LM-generated caption is present. Yet, it appears that the audio generated with an LM-caption is more likely to be identified as a cover of the original recording by CLEWS, which is trained to recognize compositional similarities between song versions \cite{serra2025supervisedcontrastivelearningweaklylabeled}.

We provide listening examples \footnote{Supplementary material: https://ismir-2026-gine.github.io/ghost-in-the-encoder/.} of songs from each genre where CLEWS correctly retrieved the original recording from its \texttt{with-CoT} generated version (R$@1$). In informal listening, it appears that songs generated \texttt{with-CoT} bear noticeably more musical and structural resemblance to the original tracks than \texttt{no-CoT}, but seemingly deviate from the expected vocal timbre, gender, or instrumentation, as a result of the auto-generated caption. Perhaps ACE-Step is reproducing covers of popular songs from its training data, in which case, a version-matching embedding like CLEWS would be invariant to these differences. As expected, the \texttt{no-CoT} songs are less varied in their style/timbre overall.

The dissociation between CoT conditions in artist-probe accuracy, cover-retrieval rate, and our exploratory perceptual observations, suggests the LM caption may function as a song-recognition trigger, surfacing training-data reproductions that lyric conditioning alone does not. We leave systematic investigation of this pathway to future work.

\begin{table}[t]
\centering
\begin{tabular}{lccc}
\toprule
$K$ & \texttt{with-CoT} & \texttt{no-CoT} & Chance \\
\midrule
10  & 0.055 & 0.022 & 0.005 \\
25  & 0.088 & 0.045 & 0.013 \\
50  & 0.137 & 0.073 & 0.025 \\
100 & 0.210 & 0.131 & 0.050 \\
200 & 0.304 & 0.204 & 0.100 \\
400 & 0.449 & 0.333 & 0.200 \\
\bottomrule
\end{tabular}
\caption{CoverID Retrieval$@K$ via CLEWS. Random baseline = $K/N$ where $N=2{,}000$.}
\label{tab:retrieval_at_k}
\end{table}

\section{Conclusion}

We demonstrate that the lyric encoder in ACE-Step 1.5 contains linear representations of artist identity that can be retrieved without any explicit artist information. These representations persist through the Diffusion Transformer's cross-attention interface and are injected into audio-token hidden states at the early layers. Using lexical classification and transformer-based embeddings as baselines, we show that this signal is acquired during pretraining rather than inherited from the input encoder, and that it is modulated by ACE-Step's CoT caption pathway in ways that warrant further study.

Our contribution highlights the potential for \textit{mechanistic} investigations of learned artist representations in music models. While our probing study suggests possible explanations for replication patterns, the potential for intervention via causal mechanisms remains to be explored.

Previous studies exposed conditioning exploits in generative models that could bypass replication safeguards by triggering memorization \cite{roh2025bobs, coelho2025artist, nagarajan2025namefree}. Our work locates a conditioning channel that can identify which artist's work the model is attempting to imitate during the generative process, which may be an interpretable form of leakage that explains these exploits.

Future interpretability work in this domain may include studying the role of style captions in exposing artist representations \cite{nagarajan2025namefree}, or extending our probing approach to other models and architectures, alongside perceptual evaluation. More broadly, we encourage others to explore latent-space auditing as a means of understanding behavioral patterns in music models at a representational level.

\section{AI Usage Statement}
Claude and Claude Code were used to help edit and grammar check the paper, clean up code, and create the supplementary website.

\section{Acknowledgements}
This work is supported by the "Cátedra IA y Música" project (TSI-100929-2023-1), funded by the Secretaría de Estado de Digitalización e Inteligencia Artificial, the European Union-Next Generation EU funds and BMAT Music Innovators. And by the "IMPA" project (PID2023-152250OB-I00) funded by MCIU/AEI/10.13039/501100011033/FEDER, UE.

\bibliography{ISMIRtemplate}

\end{document}